\documentclass[preprint,12pt]{aastex}
\usepackage{emulateapj5}
\usepackage{graphicx}

\begin{document}

\title{Supersonic Magnetic Upflows in Granular Cells Observed with Sunrise/IMaX}

\author{J.M.~Borrero$^{1,3}$, V.~Mart{\'\i}nez-Pillet$^{2}$, R.~Schlichenmaier$^{1}$, S.K.~Solanki$^{3,4}$,
J.A.~Bonet$^{2}$, J.C.~del Toro Iniesta$^{5}$, W.~Schmidt$^{1}$, P.~Barthol$^{3}$, A.~Gandorfer$^{3}$, 
V.~Domingo$^{6}$, M.~Kn\"olker$^{7}$}
\affil{$^{1}$Kiepenheuer-Institut f\"ur Sonnenphysik, Sch\"oneckstr. 6, D-79110, Freiburg, Germany\\
$^{2}$Instituto de Astrof{\'\i}sica de Canarias, Avd. V{\'\i}a L\'actea s/n, La Laguna, Spain\\
$^{3}$Max Planck Institut f\"ur Sonnensystemforschung, Max Planck Str. 2, Katlenburg-Lindau, 37191, Germany\\
$^{4}$School of Space Research, Kyung Hee University, Yongin, Gyeongg 446-701, Republic of Korea\\
$^{5}$Instituto de Astrof{\'\i}sica de Andaluc{\'\i}a (CSIC), Apdo. de Correos 3004, 18080, Granada, Spain\\
$^{6}$Image Processing Laboratory, University of Valencia, P.O. Box: 22085, E-46980 Paterna, Valencia, Spain\\
$^{7}$High Altitude Observatory (NCAR), 3080 Center Green Drive CG-1, Boulder, USA}
\email{borrero@kis.uni-freiburg.de, vmp@iac.es, schliche@kis.uni-freiburg.de, solanki@mps.mpg.de,
jab@iac.es, jti@iaa.es, wolfgang@kis.uni-freiburg.de, barthol@mps.mpg.de, gandorfer@mps.mpg.de,
Vicente.Domingo-Codonyer@uv.es, knoe@hao.ucar.edu}

\begin{abstract}
{Using the IMaX instrument on-board the {\sc Sunrise} stratospheric balloon-telescope we have detected extremely shifted polarization
signals around the \ion{Fe}{1} 5250.217 {\AA} spectral line within granules in the solar photosphere. We interpret the velocities
associated with these events as corresponding to supersonic and magnetic upflows. In addition, they are also related to the 
appearance of opposite polarities and highly inclined magnetic fields. This suggests that they are produced by the reconnection 
of emerging magnetic loops through granular upflows. The events occupy an average area of 0.046 arcsec$^2$ and last for about 80 seconds,
with larger events having longer lifetimes. These supersonic events occur at a rate of $1.3\times10^{-5}$ occurrences per second per arcsec$^{2}$.}
\end{abstract}


\shorttitle{Supersonic Magnetic Upflows seen with IMaX}
\shortauthors{Borrero et al.}
\maketitle

\def\kms{~km s$^{-1}$}
\def\deg{^{\circ}}

\section{Introduction}%

Unlike sunspots and active regions, the magnetic field in the solar granulation evolves at 
very short spatial and temporal scales as magnetic field lines are dragged and twisted by convective motions.
This has made magnetic phenomenon in the quiet Sun particularly elusive and difficult to study.
Recent instruments have achieved enough spatial resolution and temporal cadence to uncover some of 
these phenomena, such as emergence of magnetic loops: Centeno et al. (2007), Mart{\'\i}nez \& Bellot Rubio (2009),
Zhang et al. (2009), Mart{\'\i}nez et al. (2010); emergence of single polarity elements: Orozco et al. (2008); convective
collapse in intergranular lanes: Bellot Rubio et al. (2001), Nagata et al. (2008), Fischer et al. (2009);
horizontal supersonic velocities: Straus et al. (2010); supersonic downflows: Shimizu et al. (2007);
shocks: Socas-Navarro \& Manso-Sainz (2005); vortex-flows (Bonet et al. 2010), vortex-tubes (Steiner et al. 2010), etcetera.
 In this paper we report on a phenomena that has been uncovered by the IMaX instrument 
on board of the {\sc Sunrise} balloon. This phenomenon appears as highly shifted polarization signals
 within granular cells. They are usually associated with the presence of magnetic 
fields of opposing polarities in its vicinity. Highly inclined magnetic fields around these features 
are also commonly found, although not always connecting the two opposite polarities. Moreover, the 
long temporal series obtained with the IMaX instrument allow us to detect many events of similar 
characteristic and therefore to make statistics about their occurrence rate, lifetimes and sizes.

\section{Instrument and data set}%

Our data were recorded with the IMaX (Imaging Magnetrograph eXperiment; Mart{\'\i}nez Pillet et al. 2010) instrument 
on board of the {\sc Sunrise} balloon-borne observatory (Solanki et al. 2010; Barthol et al. 2010; Gandorfer et al. 2010). 
An average flight altitude of 35 km allowed {\sc Sunrise} to avoid 99 \% of the disturbances introduced by the Earth's 
atmosphere, which together with the Correlation-Tracker and Wavefront Sensor (CWS; Schmidt et al. 2004; Berkefeld et al. 2010)
and the phase diversity calibration of the PSF of the optical system and further image reconstruction yielded
spectropolarimetric data with a spatial resolution of 0.15"-0.18" and a field of view of 46"$\times$46" (after reconstruction).

Our IMaX dataset comprises two observing sequences (of 22.7 and 31.6 minutes respectively) recorded on June 9th. Both
sequences were taken close to disk center in a quiet granulation region, although the second set contains a large
network patch within the FOV. IMaX scans in 4 wavelength positions relative to $\lambda_0=$5250.217 {\AA}: $-$80, $-$40, 40, 
80 m{\AA} (hereafter denoted with the indexes $\lambda_1$,...,$\lambda_4$), across the magnetically 
sensitive ($g=3$) \ion{Fe}{1} 5250.217 {\AA} spectral line. The HWHM (half-width at half-maximum) of IMAX's transmission profile is estimated
to be 42.5 m{\AA}. This value includes the effect of the secondary peaks in the transmission profile. In addition, a fifth 
wavelength (continuum) point is measured : $\lambda_{\rm c}=\lambda_0+227$ m{\AA}. At all five wavelengths, IMaX records the 
4 polarization states of the light (Stokes $I$, $Q$, $U$ and $V$). Each full cycle (5 wavelengths and 4 polarization 
states) is taken in about 32 seconds. The noise level is of the order of 3.5$\times$10$^{-3}$ in the reconstructed
data. Each component of the Stokes vector is normalized to the average quiet Sun continuum intensity: $I_{\rm qs}$ (also
when not explicitly mentioned).

\section{Detection and Description}%

The data of Stokes $V(\lambda_{\rm c})$, or $V_{\rm c}$, shows a number of regions where the amount of circular
 polarization reaches values of 1.5-3$\times$10$^{-2}$. Hereafter, we will refer to this features as \emph{supersonic
magnetic events} due to the large velocities in a magnetized plasma required to shift the polarization 
signal so far from the line-center. Figure 1 (right panel) shows an extreme example where 4 of these features appear within a small (15"$\times$15") 
field of view\footnote{Two movies showing these events are also available on-line at: ftp://ftp.kis.uni-freiburg.de/personal/borrero/sunrise/:
{\it superson\_163\_1\_h264.mp4} and {\it superson\_163\_2\_h264.mp4}}. In our combined 54.3 minutes of observations we have detected 4441 
pixels belonging to 87 different events. This number of events yields a rate of occurrence of $1.3\times10^{-5}$
events per second per arcsec$^2$. These events have been identified as the regions in each image that 
posses $|V_{\rm c}| > 1.25\times10^{-2}$, and imposing that the patch must contain at least 9 pixels in order 
to match the instrument's spatial resolution\footnote{IMaX pixel size is 0.055", therefore, 9 pixels occupy an 
area of 0.027 arcsec$^2$ which is comparable to the instrument's 0.15$\times$0.18 arcsec$^2$ resolution}. Figure 
2a and 2b display histograms of the size (in pixels) and lifetime of these events. Here we also indicate 
with vertical dashed lines, the mean duration of these events (81.3 seconds), and their average size (15.5 
pixels or 22720 km$^2$ at disk center). Pearson's correlation coefficient between the lifetime and sizes of 
these events is 0.76, indicating that larger regions tend to live longer (Fig.~2c).

\begin{center}
\includegraphics[width=9cm]{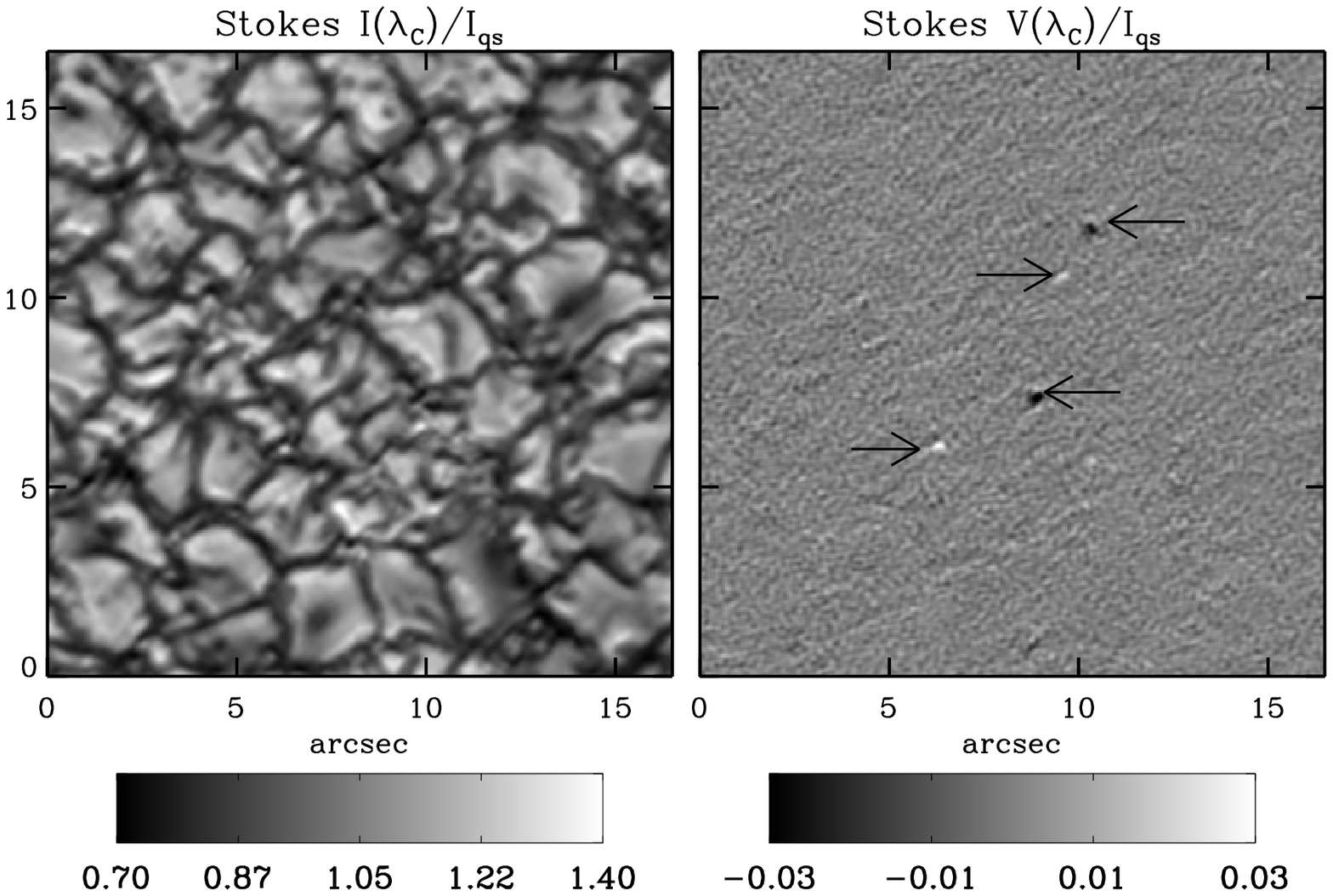}
\figcaption{{\it Left panel}: reconstructed continuum intensity map ($I_{\rm c}$) in a small
region of IMaX's field of view. {\it Right panel}: Circular polarization ($V_{\rm c}$; $+$227 m{\AA} away from line center)
map corresponding to the same region as the left panel. Indicated with arrows are 4 patches, where the circular
polarization (absolute value) reaches values larger than 0.0125 \%.}
\end{center}

Histograms in Fig.~2a and 2b peak at the lower limit of our detection thresholds: 10 pixels in size 
and 32 seconds in lifetime. This suggests that there could be undetected events at smaller spatial 
scales and with shorter lifetimes. Although a minority, there are also some instances of events that 
embody areas larger than 25 pixels (10 \% of the total) or last more than 3 minutes (11.5 \% of the total).

Figure~3 displays $I_{\rm c}$ (upper row), $V_{\rm c}$ (middle row), line-of-sight velocity $v_{\rm LOS}$ (bottom row), 
whereas Figure~4 displays the line-average circular polarization: $\frac{1}{4}\sum_{i=1}^4 |V_i|$  (signed with the 
sign of $V_1$; upper row), and line-average linear polarization: $\frac{1}{4}\sum_{i=1}^4 \sqrt{Q_i^2+U_i^2}$ (lower row) 
for three selected \emph{supersonic events}. It is important to mention that the line-of-sight velocity in Fig.~3 has been 
obtained from a Gaussian fit to the first four wavelength positions in Stokes $I$, therefore it should not be surprising 
that we do not see \emph{supersonic events} here, but rather in $V_{\rm c}$ (second row) because this is where we have 
defined them. Also, the velocities presented here have already been corrected for the Sun's gravitational red-shift, 
convective blue-shift, and shifts caused by the collimated beam configuration of the instrument.

All three samples in Fig.~3-4 occur at the center of granular cells: in regions with large blue-shifted Doppler velocities and large 
continuum intensities. In fact, 72 \% of all detected events occur at the center or edges of granules, whereas only 8 \% appear 
in intergranular lanes. The remaining 20 \% occur in evolving granulation (i.e: on a granule that turns into an intergranule or 
vice-versa) and might be related to exploring granules (Rast 1995). 

In addition, the examples in Figs.~3 and 4 show that sign of $V_{\rm c}$ bears no relation to the polarity of the magnetic field 
as $V_{\rm c}$ can have the same or opposite sign as $V_1$. In addition, most of the detected supersonic events appear as a single 
patch of large $|V_{\rm c}|$, although sometimes (e.g: first event in Figs.~3-4), specially in regions where the magnetic 
field topology is very complex, two supersonic patches appear together.

When looking at the vicinity of these events (field of view of 2" around each) we observe that about 22 \% are 
associated with one single magnetic polarity in their neighborhood. In this case, the enhanced $V_{\rm c}$ occurs always at precisely 
the same location as this single polarity patch. The remaining 70 \% of all detected events are related to the appearance of 
opposite magnetic polarities within 2" of the supersonic events. Moreover, whenever opposite polarity regions are seen in the vicinity of 
the supersonic events, patches of enhanced linear polarization (not always connecting the opposing polarities) are observed
(see also Danilovic et al. 2010). As seen in the line-averaged circular and line-averaged linear polarization (Fig.~4) 
all three examples fall into this category. Although the levels of linear polarization (around 1 \%) are smaller than those 
of circular polarization (2-3 \%), the fact that Stokes $Q$ and $U$ are clearly detected above the noise level is already evidence 
for the presence of highly inclined magnetic fields\footnote{The relation between the the appearance of \emph{supersonic} events
and the existence of horizontal fields in their surroundings (in 70\% of all cases) is showcased in the online movies located at
ftp://ftp.kis.uni-freiburg.de/personal/borrero/sunrise/: {\it Vc$+$linpol\_1\_h264.mp4} and {\it Vc$+$linpol\_1\_h264.mp4}}.
Another feature of the \emph{supersonic events} is that they do not stay always at the same position on the 
solar surface, but rather, they move horizontally with velocities typical of the granulation ($1-2$\kms; cf. 
Straus et al. 2010).

\begin{figure*}
\begin{center}
\includegraphics[width=17cm]{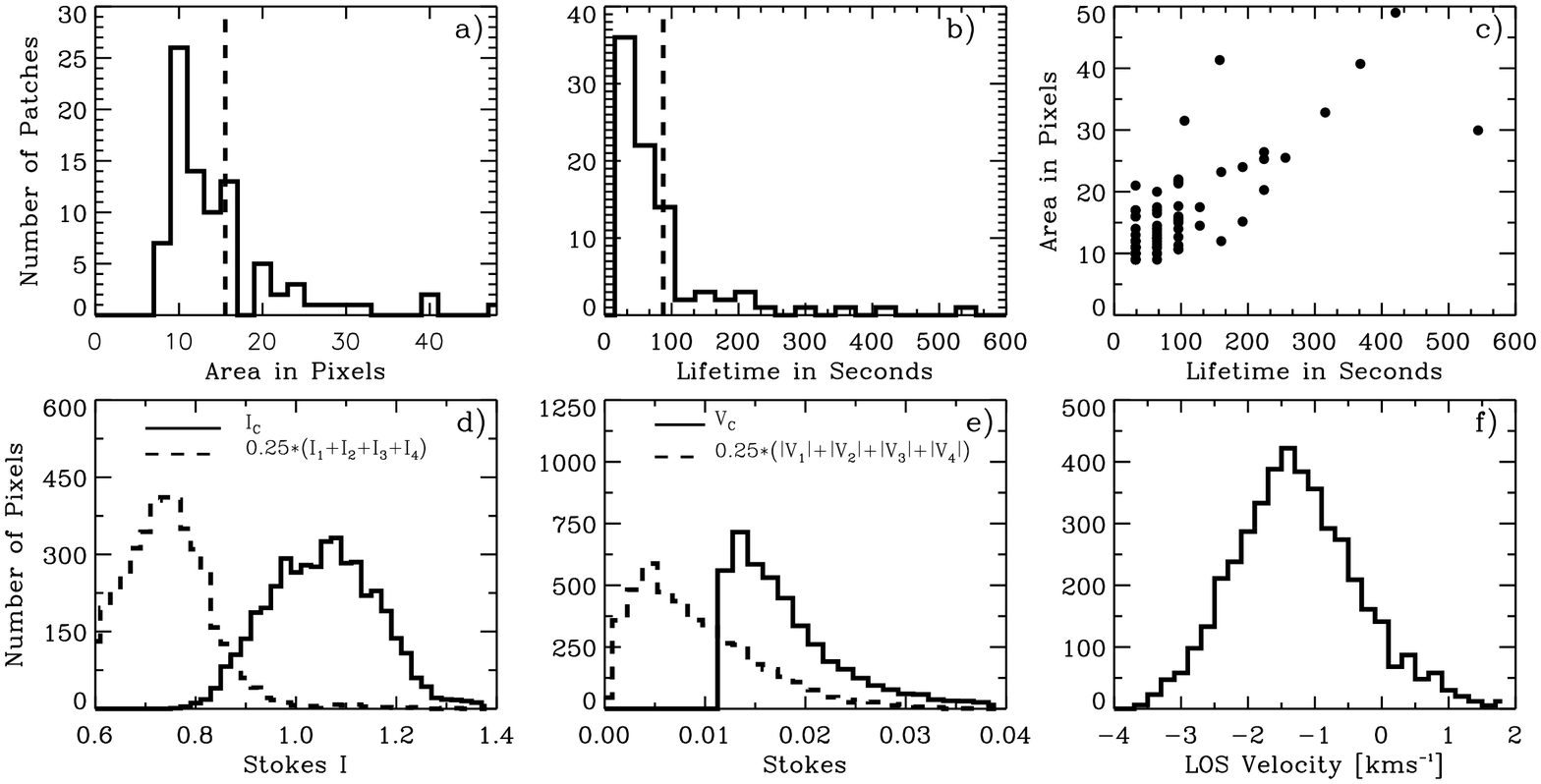}
\figcaption{{\it a)} Histograms of event area in number of pixels. {\it b)} Histogram of events' lifetime. 
The vertical dashed line in the first two panels indicates the mean value of the size and lifetime, respectively. {\it c)}
Surface area as a function of lifetime for the 87 detected supersonic events. {\it d)} Histogram of 
the number of pixels having a given line-average intensity (dashed) and a given continuum
intensity $I_{\rm c}$ (solid). {\it e)} Histogram of the number of pixels having a given line-average 
circular polarization (dashed) and a given continuum circular polarization $V_{\rm c}$ (solid). {\it f)}
Histogram for the line-of-sight velocity, obtained from the first four wavelengths in Stokes $I$,
in the pixels displaying large values of $V_{\rm c}$.}
\end{center}
\end{figure*}

\begin{figure*}
\begin{center}
\includegraphics[width=12cm]{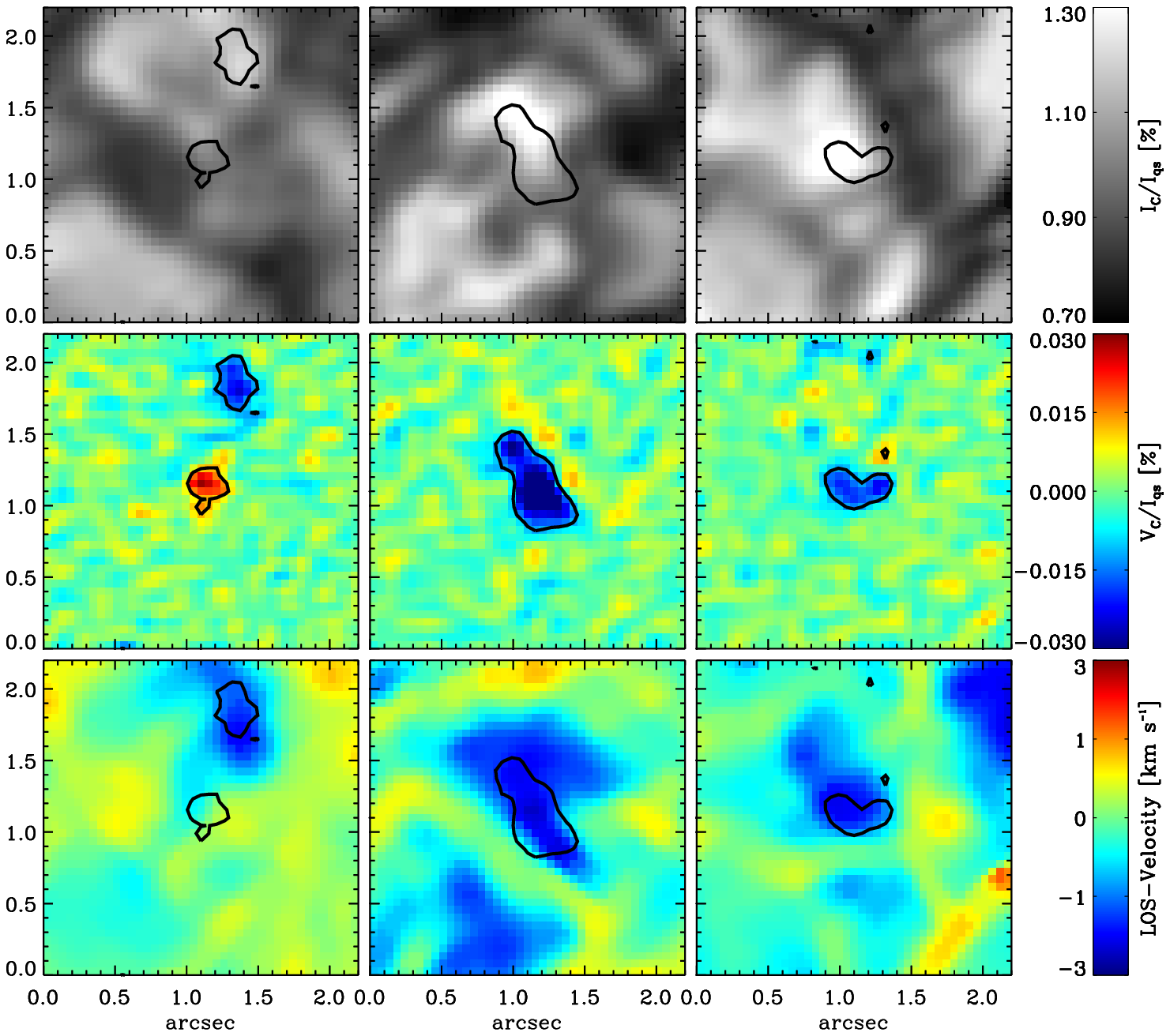}
\figcaption{{\it From top to bottom}: Continuum intensity $I_{\rm c}$ (normalized to the quiet Sun average $I_{\rm qs}$), 
continuum circular polarization $V_{\rm c}$ (normalized to the quiet Sun continuum intensity $I_{\rm qs}$), and 
line-of-sight velocity $v_{\rm LOS}$ (Gaussian fit to first four wavelengths of Stokes $I$). {\it From left to right}: 
three different examples of supersonic events found.}
\end{center}
\end{figure*}

The fact that most of the detected events are associated with the presence of magnetic fields of opposite
polarity and inclined magnetic fields in their vicinity indicates that they are likely to be related to some form of 
magnetic reconnection. In addition (see Fig.~2f), the supersonic events occur mostly in blue-shifted regions, 
suggesting that the magnetic field emerges as it is dragged upwards by upflowing granules. The same conclusion 
can be reached by looking at the solid line in Fig.~2d. This plot shows that the continuum 
intensity $I_{\rm c}$ is in most cases larger than 1.0, meaning that the supersonic events occur mainly in the brighter regions of the 
granulation (upflowing granules). A possible interpretation of these observations is that the supersonic magnetic flows  
are associated with the emergence of magnetic loops (where the field is horizontal) that reconnect with the preexisting ambient magnetic field. 
The footpoints on this loop would be seen as the opposite polarities that we observe in many of the cases. 

\begin{figure*}
\begin{center}
\includegraphics[width=12.2cm]{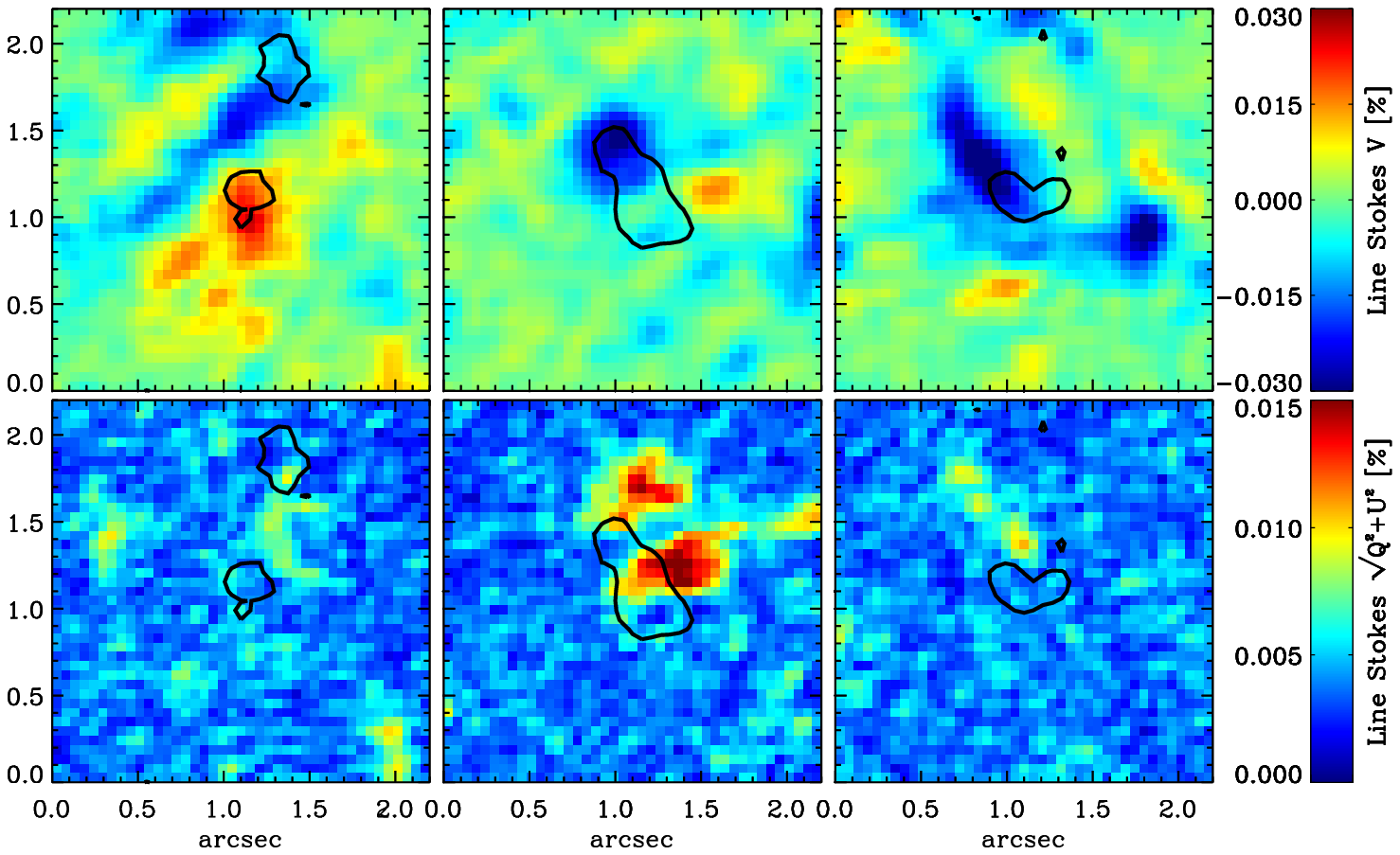}
\figcaption{Same as Fig.~3 but for the signed line-average circular polarization (top) and line-average
linear polarization (bottom).}
\end{center}
\end{figure*}

\section{Discussion and Conclusions}%

Since the line-of-sight velocities associated to these events are mostly blue-shifted, we surmise that the 
signal we observe in $V_{\rm c}$ (+227 m{\AA} on the red side of $\lambda_0=5250.217$ {\AA}; see Fig.~1) 
is not due to a red-shifted contribution from \ion{Fe}{1} 5250.217 {\AA}, but rather a blue-shifted contribution 
from the nearby line \ion{Fe}{1} 5250.653 {\AA}\footnote{We have conducted a spectral line search in NIST 
(http://physics.nist.gov/PhysRefData/ASD/lines$_{}$form.html) and VALD databases (Piskunov et al. 1995) 
and found only this other spectral line in the vicinity of Fe I 5250.217 {\AA}.}. Note that the continuum
wavelength $\lambda_{\rm c}$ is almost as far from \ion{Fe}{1} 5250.217 as from \ion{Fe}{1} 5250.653.

In principle, a wavelength difference of 227 m{\AA} corresponds to a Doppler shift of 12\kms. However, 
a better estimation that considers the widths of the spectral line and IMAX's transmission profile, yields 
a possible range of $v_{\rm LOS} \in [5,12]$. Because of projection effects, $v_{\rm LOS}$ represents only 
a lower limit of the absolute velocity. Taking into account that the speed of sound in the solar photosphere 
is about $c_s \simeq 6$\kms, this points towards potentially supersonic line-of-sight velocities. 
With the current data, however, we do not sample the spectral region well enough to narrow down further
the range of possible velocities.

One problem facing an interpretation in terms of supersonic upflows is that, if \ion{Fe}{1} 5250.653 {\AA}
was affected by 12\kms upward velocities, we would observe (along with large $V_{\rm c}$ signal) a 
decrease in the $I_{\rm c}$ signal, since the absorption line would have also shifted from 5250.653 {\AA}
into $\lambda_{\rm c}$. However, Fig.~2d indicates that the intensity observed at $\lambda_{\rm c}$ (solid line) is much 
larger than the mean intensity observed around the spectral line Fe I 5250.217 {\AA} (dashed line), suggesting 
that not the whole line is shifted. On the other hand, smaller velocities ($v_{\rm LOS} \approx -4$\kms) do not 
help solve the puzzle either, since then the line-average circular polarization in Fe I 5250.217 {\AA} 
would still be rather large, in particular, much larger than the circular polarization in the continuum $V_{\rm c}$. 
However, as shown in Fig.~2e, $V_{\rm c}$ (solid line) is larger than the line-average circular polarization (dashed line).

We have not studied in detail additional complications such as strong gradients along
the line-of-sight or the presence of unresolved structures within the area occupied by the 
supersonic event, which produce peculiar Stokes profiles and ought to be analyzed in a
different way. Note however, that in both of these cases, extremely large velocities at some height in 
the atmosphere or within a small portion of the resolution element, must still be invoked (see for
example Fig.~4 in Socas-Navarro \& Manso-Sainz 2005). An additional 
effect that we have not considered here, but which may be important to explain the seemingly contradictory
behavior described in the previous paragraph, is the fact that, if indeed these events are related to magnetic 
reconnection (as supported by the presence of opposite magnetic polarities), one or both spectral lines, 
could appear in emission, making it possible to have large blue-shifted velocities without noting a decrease 
in $I_{\rm c}$, but rather an increase.

\begin{acknowledgements}
The German contribution to {\sc Sunrise} is funded by the Bundesministerium
f\"{u}r Wirtschaft und Technologie through Deutsches Zentrum f\"{u}r Luft-und Raumfahrt e.V. 
(DLR), Grant No. 50~OU~0401, and by the Innovationsfond of the President of the Max Planck 
Society (MPG). The Spanish contribution has been funded by the Spanish MICINN under projects 
ESP2006-13030-C06 and AYA2009-14105-C06 (including European FEDER funds). The HAO
contribution was partly funded through NASA grant number NNX08AH38G. This work has been partly
funded by the WCU grant No. R31-10016 funded by the Korean Ministry of Education, Science \&
Technology.
\end{acknowledgements}

\end{document}